\documentstyle[12pt,epsf]{article}

\begin{document}

\author{ B. Boisseau\thanks{E-mail : boisseau@celfi.phys.univ-tours.fr},
C. Charmousis\thanks{E-mail : christos@celfi.phys.univ-tours.fr}
and B. Linet\thanks{E-mail : linet@celfi.phys.univ-tours.fr} \\
\small Laboratoire de Math\'ematiques et Physique Th\'eorique \\
\small CNRS/EP 93, Universit\'e Fran\c{c}ois Rabelais \\
\small Facult\'e des Sciences et Techniques \\
\small Parc de Grandmont 37200 TOURS, France }

\title{\bf Dynamics of a self-gravitating \\ thin cosmic string}

\date{}
\maketitle

\begin{abstract}
 
We assume that a self-gravitating  thin string can be locally described by what we shall call a {\it smoothed cone}. If we impose a specific constraint on the model of the string, then its central line obeys the Nambu-Goto equations. If no constraint is added, then the worldsheet of the central line is a totally geodesic surface. 

{\em PCAS numbers : 04.20.-q, 11.27.+d}
\end{abstract}

\thispagestyle{empty}

\newpage
\renewcommand{\thesection}{\Roman{section}}
\section{Introduction}

In the framework of general relativity, two theoretical aspects of 
cosmic strings have been mainly studied. The dynamical one in which
the equations of motion of the infinitely thin
cosmic string are governed by the Nambu-Goto action and secondly the 
self-gravitating one 
in which the straight cosmic string in particular is considered 
as the source of a gravitational field.
In this case one finds that the asymptotical metric generated 
by the straight cosmic string is a conical metric \cite{she} and 
the case of a singular line is obtained in a limit process.

The study of line sources in gravitation can be originated from a 
paper of Israel \cite{isr1} in which the author concluded that there existed 
"no simple general prescription [...] for obtaining
the physical characteristics of an arbitrary line source". However, Vilenkin 
\cite{vil} gave the physical meaning of the conical singularity 
as being a cosmic string. Attention has been
recently devoted in understanding the dynamics of a conical-type line
source of the Einstein equations \cite{vic,fro,unr,cla}. The conclusions 
summarised in \cite{isr2} are as follows.
A two-dimensional timelike worldsheet whose points are conical 
singularities of the 4-geometry cannot be in general the dynamical 
evolution of a Nambu-Goto string of arbitrary initial shape. 
The conical singularity requires that the worldsheet is totally geodesic. 
This restraints the initial shape of the string as well as the evolution 
of each of its points to be a geodesic of the 4-geometry.

In this paper, we take up again the question of the dynamics but for
a self-gravitating extended string. For a straight cosmic string, it 
is known that the exterior metric may be matched with an interior metric
having as source an energy-momentum tensor with suitable properties.
We consider a string of arbitrary shape but we restrict ourselves
 to the case of
a thin tube of matter adopting the assumption that the exterior metric is
locally the one describing a straight cosmic string. 
The interior metric is constrained from this and
the energy-momentum tensor could be derived from the metric. The aim
of the present work is to find the equations of motion
of the central line of this thin tube of matter in the limit where its
radius tends to zero.

We emphasize that we do not consider directly a self-gravitating line.
In our method, the conical points are smoothed out on a scale
comparable to the radius of the string.
The central line of this tube sweeps a timelike 2-worldsheet whose 
points are perfectly regular. A local coordinate system can then be 
attached to our spacetime by taking the two parameters of the
 worldsheet 
as the first two coordinates and the other two as geodesic
 coordinates 
pointing in a direction orthogonal to the worldsheet. This local
 coordinate 
system affixed to the worldsheet allows the natural
 introduction of the 
extrinsic curvature and other geometric parameters of the worldsheet.

The interior metric of the string is essentially 
characterised by a function 
$f(l)=\epsilon h(l/\epsilon)$ in which $l$ is the radial coordinate whose 
origin lies on the worldsheet and $\epsilon$ is a length typical of the 
thickness of the string. The function $h$ is arbitrary, only submitted to 
certain conditions expressing the smoothness of the spacetime on the 
worldsheet, as well as the matching conditions on the boundary between the 
interior of the string and the vacuum.

If no other conditions are imposed, we shall prove that 
the worldsheet tends to a totally geodesic surface, {\em i.e.} of
vanishing extrinsic curvature, when $\epsilon$ goes to zero. 
However we have found another possibility. If we impose on the function 
$h$ a specific supplementary constraint then the extrinsic curvature no 
longer tends to zero.
Nevertheless the mean curvature tends to zero when $\epsilon$ does and thus 
the worldsheet tends to be locally extremal which is precisely the behavior 
of the Nambu-Goto string. Since the function $h$ is a characteristic 
of the metric in the interior of the string  thus the constraint imposed 
is interpreted as a specification of the matter of the string.

The paper is organised in the following manner. We recall in Sec. II the basic results on self-gravitating straight strings with some 
thickness and we describe the formalism that we shall use. In Sec. III, a  self-gravitating  string of arbitrary shape is 
introduced and the coordinate system adopted to its study is introduced. 
In Sec. IV, we expand in powers of $1/\epsilon$ the geometrical 
quantities appearing in the Einstein equations. In Sec. V, taking the 
limit of $\epsilon$ going to zero of the Einstein equations on the boundary 
between the string and the vacuum, we obtain the constraints that the 
worldsheet swept by the central line of the string must satisfy. 
Finally in Sec. VI, we find the Nambu-Goto energy-momentum tensor in the 
zero limit of $\epsilon$, confirming the choice of the definitions adopted
in Sec. III.  
   
\section{Straight strings as smoothed cones}

In a general relativistic context, it is possible to portray a straight 
string as a thin cylinder of matter \cite{mar,got,his,lin}. We give some basic 
features of a self-gravitating straight string since they will be used 
later on in the general case of strings of arbitrary shape.

In the coordinate system $(t,z,l,\phi)$ with $l\geq0$ and 
$0\leq\phi<2\pi$, the energy-momentum tensor has the form
\begin{equation}
\label{1}  
T^{t}_{t}=T^{z}_{z}=-\sigma(l)\qquad T^{l}_{l}=T^{\phi}_{\phi}=0
\qquad 0\leq l<l_{\small{0}}
\end{equation}
where $l_{\small{0}}$ is the given radius of the cylinder. The energy 
density $\sigma$ is a  positive regular function of  $\Re^{2}$. Taking 
into account the Einstein equations, the interior metric can be written as
\begin{equation}
\label{2}
ds_{\small{INT}}^{2}=-dt^{2}+dz^{2}+dl^{2}+f^{2}(l)d\phi^{2}, \qquad  
0\leq l<l_{\small{0}}
\end{equation}
where the positive function $f$ determines $\sigma$ by the formula,
\begin{equation}
\label{3}
\sigma=- \frac{1}{8\pi G}\frac{f^{\prime\prime}}{f},
\end{equation}
$G$ being the Newtonian gravitational constant.
In order to ensure a regular behavior of metric (\ref{2}) at $l=0$, 
one must choose $f$ such that
\begin{equation}
\label{4}
f(l)\sim l+a_{3} l^{3}+a_{5} l^{5}+\cdots  \qquad
 \mbox{as} \qquad l\rightarrow 0.
\end{equation}
We can also require that the function $f$ is increasing.
The exterior metric which can be matched to the interior metric (\ref{2}) 
can be expressed in the form 
\begin{equation}
\label{5}
ds_{\small{EXT}}^{2}=-dt^{2}+dz^{2}+dl^{2}+\sin^{2}\alpha\, 
(l-\stackrel{-}{l_{\small{0}}})^{2}d\phi^{2},
\qquad l>l_{\small{0}}
\end{equation}
where the constants $\alpha$ and $\stackrel{-}{l_{\small{0}}}$ are determined 
from the following matching conditions of the metric and its first
derivatives at $l=l_{0}$
\begin{equation}
\label{6}
f(l_{\small{0}})=\sin\alpha(l_{\small{0}}-\stackrel{-}{l_{\small{0}}})
\quad {\rm and} \quad
f^{\prime}(l_{\small{0}})=\sin\alpha.
\end{equation}

The linear mass density $\mu$ of the straight string is defined by
\begin{equation}
\label{8}
\mu=\int_{l<l_{\small{0}}}\sigma f\, dl\, d\phi.
\end{equation}
By using (\ref{3}), (\ref{4}) and (\ref{6}), it is easy to see that it
is related to the angular deficit $\Delta$ of metric (\ref{5}) by
\begin{equation}
\label{9}
\Delta=2\pi (1-\sin\alpha)=8\pi G \mu.
\end{equation}
From (\ref{9}),
we note that the exterior metric (\ref{5}) is independent of the 
details of the internal string as well as of its radius.

At this point, it is enlightening to recall the solution for a straight 
string with $\sigma=$constant.
The interior metric is characterised by 
\begin{equation}
\label{10}
f(l)=\epsilon\sin\frac{l}{\epsilon}\qquad\mbox{and}\qquad\sigma=
\frac{1}{8\pi G\epsilon^{2}}
\end{equation}
and the exterior metric is again (\ref{5}). The geometrical
interpretation of the parameter $\epsilon$ will be
given below. The matching conditions 
(\ref{6}) give the relation $l_{\small{0}}/\epsilon =\pi /2 -\alpha$ 
and thereby
\begin{equation}
\label{11}
\epsilon=(l_{\small{0}}-\stackrel{-}{l_{\small{0}}})
\frac{\sin\alpha}{\cos\alpha}.
\end{equation}
We point out that $l_{\small{0}}/\epsilon$ depends only on $\alpha$.

Returning to the general case this particular solution suggests to 
impose in the generic interior metric (\ref{2}) the following form
\begin{equation}
\label{12}
f(l)=\epsilon h(\frac{l}{\epsilon})
\end{equation}
where $h$ is a smooth function and $\epsilon$ is a
parameter which takes again the value (\ref{11}).
Now the matching conditions (\ref{6}) yield simply
\begin{equation}
\label{13}
h(\frac{l_{\small{0}}}{\epsilon})=\cos\alpha
\quad {\rm and} \quad
h^{\prime}(\frac{l_{\small{0}}}{\epsilon})=\sin\alpha.  
\end{equation}
For a given function $h$, the quotient $l_{0}/\epsilon$ depends only on the constant angle
$\alpha$, {\em i.e.} on the linear mass density $\mu$ 
by virtue of relation (\ref{9}).

One can give a geometrical interpretation. By performing the change of 
coordinate $\rho=l-\stackrel{-}{l_{\small{0}}}$ in metric (\ref{5}), 
we recognise the so called conical metric representing a cone of 
half angle $\alpha$.
It is usefull to give a representation in a Euclidean 3-space of the 
2-surface $t=$ constant and $z=$ constant for the interior and exterior 
metrics (cf. fig. 1). It can be visualised as a cone whose top is cut out at a 
distance  $\rho_{\small{0}}=l_{\small{0}}-\stackrel{-}{l_{\small{0}}}$ 
of the vortex and replaced by an axisymmetric cap which joins the cone 
tangently along the circle of junction of radius 
$\rho_{0}\sin \alpha$.
The coordinate $l$ represents the length of the radial geodesic of the 
2-surface originated at the top of the cap. It takes the value 
$l_{\small{0}}$ at the boundary between the smooth cap and the cone
 and then it continues along a generatrix of the cone. 
The coordinate $\phi$ is the azimutal angle. Referring to fig. 1,
 we see that
$\epsilon =\rho_{0}\sin \alpha /\cos \alpha$, 
which represents the distance of the junction to the central axis, coincides with choice (\ref{11}). For the particular solution 
(\ref{10}) we have a sperical cap of radius $\epsilon$.
\begin{figure}[htbp]
\epsfxsize=14cm
\epsfysize=14cm
$$
\epsfbox{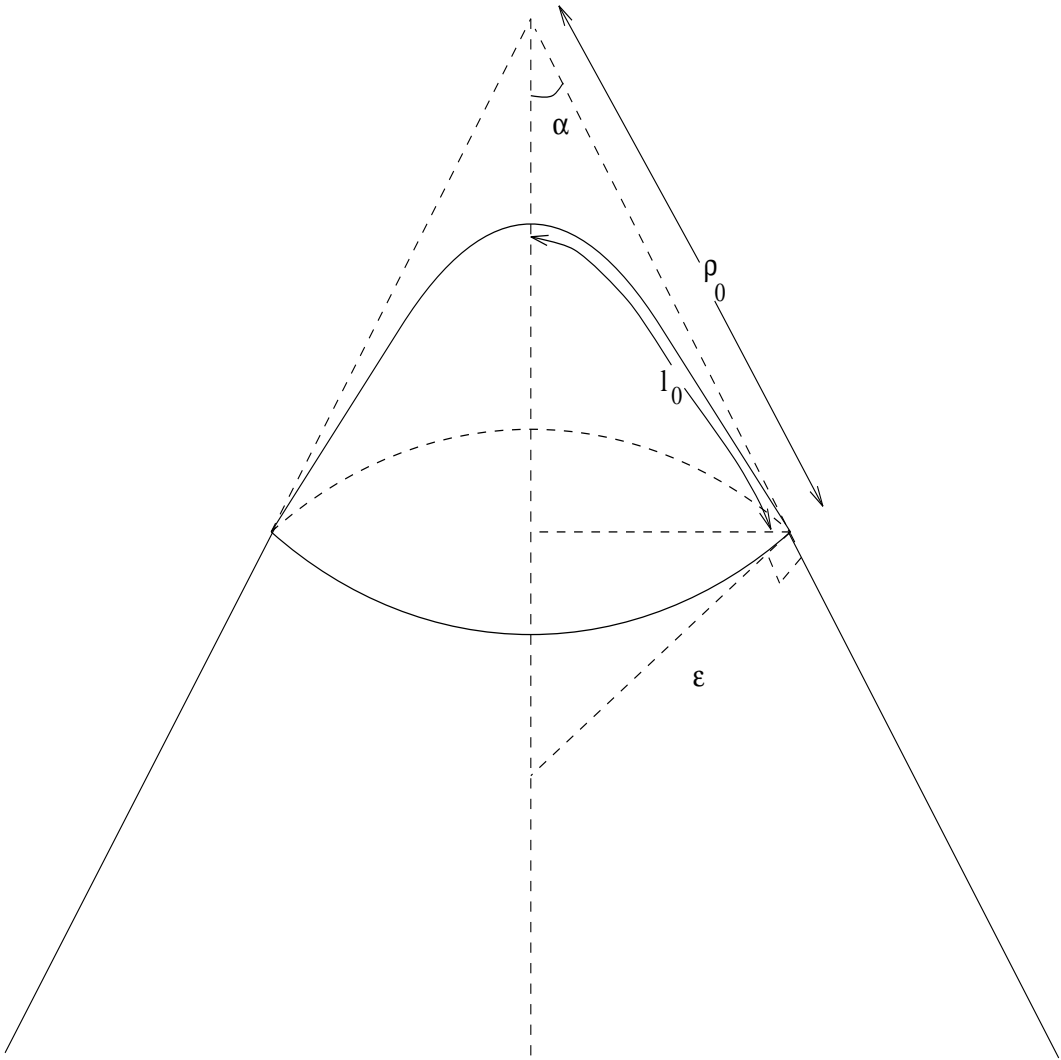}
$$
\caption{Smoothed cone}
\label{fig:Smoothed cone}
\end{figure}

In the generic case, $h$ is characteristic of the form of the cap and
$\epsilon$ of its size. This explains the choice of form (\ref{12})
of $f$ which allows us to wedge in the cone a cap of a given form
of any size.
In short, the 2-surface of  the Euclidean 3-space will be called a 
{\it smoothed cone} of half angle $\alpha$. 
The corresponding spacetime will be said to have a {\it smooth conical
 point}. We emphasize that 
 we can choose $l_{\small{0}}$ and $\epsilon$ as little as we 
want thus making the smooth cone as sharp as we want, but keeping 
the quotient  $l_{\small{0}}/\epsilon$ fixed.

Finally, we shall also require the continuity of the second derivatives of 
the metric at the junction. This is simply achieved by putting
\begin{equation}
\label{16}
f^{\prime\prime}(l_{\small{0}})=0\qquad \mbox{or} \qquad 
h^{\prime\prime}(\frac{l_{\small{0}}}{\epsilon})=0.
\end{equation}
As a consequence the Ricci tensor is continuous at the junction.
 Let us note 
that the supplementary condition (\ref{16}) is not verified by the 
particular solution (\ref{10}). This was obvious from the beginning 
since the energy density $\sigma$ is constant in the interior of a 
straight string and decreases sharply to zero at the boundary. 
The supplementary condition (\ref{16}) is physically natural for a 
thick relativistic cosmic string but not absolutely necessary. 
Its real usefulness is that the continuity of the Ricci tensor makes easier 
further calculations in the neighborhood of the boundary between 
the cap and the cone.

\section{Self-gravitating string of arbitrary shape and coordinate system}

In the previous section we described a straight self-gravitating string 
with some thickness.
The aim of this section is to extend this construction to a string of 
arbitrary shape.
If the string thickness is sufficiently small and if the central line of 
the string spans a sufficiently smooth world sheet, we can suppose 
that the   
string approaches a straight one and thus it can be locally characterised 
by a smooth conical point.
This suggests to determine the spacetime in the neighborhood of a small 
portion of the string in the following way : \\
- the interior metric given by
\begin{equation}
\label{17}
ds_{\small{INT}}^{2}=g_{AB}(\tau^{A},l,\phi)d\tau^{A}d\tau^{B}+dl^{2}+
f^{2}(l)d\phi^{2} \qquad  0\leq l\leq l_{\small{0}}
\end{equation}
where $f(l)=\epsilon h(l/\epsilon)$ was specified in Sec. II and
the metric components  $g_{AB}$ ($A,B=0,3$) are smooth. \\
- in the vicinity of the string, the exterior metric given by
\begin{equation}
\label{18}
ds_{\small{EXT}}^{2}=g_{AB}(\tau^{A},l,\phi)d\tau^{A}d\tau^{B}+dl^{2}+
\sin^{2}\alpha\, (l-\stackrel{-}{l_{\small{0}}})^{2}
d\phi^{2}\qquad l>l_{\small{0}}.
\end{equation}
We note that {\em a priori} we have ommited cross terms in (\ref{17}) and 
(\ref{18}) in order to simplify the calculus.
Metric (\ref{17}) is regular but the coordinate system  breaks down 
at $l=0$. 
We introduce coordinates $\rho^{a}$ $(a=1,2)$ which are well defined 
\begin{equation}
\label{19}
\rho^{1}=l\cos\phi\qquad {\rm and} \qquad\rho^{2}=l\sin\phi.
\end{equation}
Then, metrics (\ref{17}) and (\ref{18}) become
\begin{equation}
\label{20}
ds^{2}=g_{AB}(\tau^{A},\rho^{a})d\tau^{A}d\tau^{B}+
g_{ab}(\rho^{a})d\rho^{a}d\rho^{b}.  
\end{equation}

Let us introduce the function
\begin{equation}
\label{21}
r(l)=\frac{f^{2}(l)}{l^{4}}-\frac{1}{l^2} \quad {\rm with} \;
l=\sqrt{(\rho^{1})^{2}+(\rho^{2})^{2}}
\end{equation}
 which is well defined and smooth in the interval $0\leq l\leq l_{\small{0}}$
from (\ref{4}). It has null odd derivatives at $l=0$.
The components $g_{ab}$ have the simple expression
\begin{equation}
\label{22}
g_{11}=1+r(l)(\rho^{2})^{2},\qquad g_{12}=-r(l)\rho^{1}\rho^{2},\qquad 
g_{22}=1+r(l)(\rho^{1})^{2}
\end{equation}
in the interval 
$0\leq l \leq l_{\small{0}}$.
It is interesting to note that
$g_{ab}=\delta_{ab}+O(\rho^{2})$. There will be no need in the following
to express explicitely $g_{ab}$ for $l>l_{0}$.

The central line of the string defined by $l=0$ spans a timelike worldsheet 
parametrised by $\tau^{A}$ whose induced metric is simply 
\begin{equation}
\label{24}
\gamma_{AB}(\tau^{A})=g_{AB}(\tau^{A},\rho^{a}=0).
\end{equation}
A radial geodesic on the 2-surface $\tau^{A}=$ constant, is also a geodesic 
of the spacetime, normal to the worldsheet at the point $P(\tau^{A})$.
The coordinate  
$l$ represents the length along this geodesic measured from the point  
$P(\tau^{A})$. So, the 2-surface  $\tau^{A}=$ constant is generated by the spacetime geodesics  
tangent at $P(\tau^{A})$ to a 2-plane orthogonal to the worldsheet.
Hence we recognise in the coordinate system $(\tau^{A},\rho^{a})$ a known 
system (see for example \cite{boi}) where $\rho^{a}$ are geodesic coordinates and 
from which it is easy to extract the extrinsic curvature $K_{aAB}$ of the 
worldsheet by expanding the metric components
\begin{equation}
\label{25}
g_{AB}(\tau^{A},\rho^{a})=\gamma_{AB}(\tau^{A})
+2K_{aAB}(\tau^{A})\rho^{a}+O(\rho^{2}).
\end{equation}
As we shall see in Sec. V,  the extrinsic curvature is implicated 
in the dynamics of a self-gravitating string.

\section{Expansion of geometrical quantities}

The aim of this section is to expand in powers of 1/$\epsilon$ 
geometrical quantities such as the metric, connection and Ricci tensor 
when $\epsilon$, that is a measure of the thickness of the string, is small 
but not null. Let us point out that all length parameters 
$\rho_{\small{0}}^{a},l_{\small{0}},\epsilon$ are of the same order since $l_{\small{0}}/\epsilon$ 
is constant.

In the interval  $0\leq l\leq l_{\small{0}}$ all these quantities depend on 
the function $r(l)$ through the metric components $g_{ab}(\rho^{a})$. 
However since 
$r(l_{\small{0}})=O(1/\epsilon^{2})$ it is preferable
to substitute $r(l)$ by a function of $l/\epsilon$
\begin{equation}
\label{26}
q( \frac{l}{\epsilon})=\epsilon^2 r(l)=\frac{\epsilon^{4}}{l^{4}}h^{2}( 
\frac{l}{\epsilon})-\frac{\epsilon^{2}}{l^{2}}.
\end{equation}
It is evident that $q(l_{\small{0}}/\epsilon)$ is fixed and 
depends only of the angle $\alpha$. The same holds for its first and second derivatives $q^{\prime}(l_{\small{0}}/\epsilon)$ and
$q^{\prime\prime}(l_{\small{0}}/\epsilon)$.

The metric components (\ref{22}), their inverse, and the determinant 
can now be written  
\begin{equation}
\label{27}
g_{ab}=\delta_{ab}+q(\frac{l}{\epsilon})\epsilon_{a}^{\mbox{ }c}
\epsilon_{b}^{\mbox{ }d}\frac{\rho_{c}\rho_{d}}{\epsilon^{2}},
\end{equation}
\begin{equation}
\label{28}
g^{ab}=\frac{\delta^{ab}+q(\frac{l}{\epsilon})\frac{\rho^{a}\rho^{b}}
{\epsilon^{2}}}{1+\frac{l^{2}}{\epsilon^{2}}q(\frac{l}{\epsilon})},
\end{equation}
\begin{equation}
\label{29}
\stackrel{\wedge}{g}=det(g_{ab})=1+\frac{l^{2}}{\epsilon^{2}}
q(\frac{l}{\epsilon})
\end{equation}
where $\epsilon_{a}^{\mbox{ }c}$ is the totally antisymmetric Levi-Civita 
symbol and $\stackrel{\wedge}{(\mbox{ } )}$ stands for the induced geometrical quantities of the 2-dimensional smoothed cone.

Then one immediately gets
\begin{equation}
\label{30}
(g_{ab})_{\small{l=l_{\small{0}}}}=O(1),\qquad (g^{ab})
_{\small{l=l_{\small{0}}}}=O(1),\qquad 
(\stackrel{\wedge}{g})_{\small{l=l_{\small{0}}}}=O(1).
\end{equation}
By direct calculation of the first derivative one obtains 
\begin{equation}
\label{31}
(g_{ab,c})_{\small{l=l_{\small{0}}}}=O(\frac{1}{\epsilon}),\qquad 
(g_{,\mbox{ }c}^{ab})_{\small{l=l_{\small{0}}}}=O(\frac{1}{\epsilon}).
\end{equation}
It is also easy to see that 
\begin{equation}
\label{32}
(g_{AB})_{\small{l=l_{\small{0}}}}=O(1), \\
\qquad (g_{AB,d})_{\small{l=l_{\small{0}}}}=O(1),\qquad (g_{AB,C}
)_{\small{l=l_{\small{0}}}}=O(1).
\end{equation}
 
Now we can calculate the order of the non null connection coefficients 
for $l=l_{\small{0}}$
\begin{equation}
\label{33}
\Gamma_{BC}^{A}=\frac{1}{2}g^{AD}(g_{DB,C}+g_{DC,B}-g_{BC,D})=O(1),
\end{equation}
\begin{equation}
\label{34}
\Gamma_{aB}^{D}=\frac{1}{2}g^{AD}g_{AB,a}=O(1),
\end{equation}
\begin{equation}
\label{35}
\Gamma_{AB}^{a}=-\frac{1}{2}g^{ab}g_{AB,b}=O(1),
\end{equation}
\begin{equation}
\label{36}
\Gamma_{ab}^{c}=\frac{1}{2}g^{cd}(g_{da,b}+g_{db,a}-g_{ab,d})=
\stackrel{\wedge}{\Gamma}_{ab}^{c}(\rho^{a})=O(\frac{1}{\epsilon}).
\end{equation}
We can now deal with the Ricci tensor for $l=l_{\small{0}}$. 
The derivatives of (\ref{35}) will 
give terms of the order of $1/\epsilon$, whereas the derivatives of 
(\ref{31}) and (\ref{36}) will give terms of the order  $1/\epsilon^{2}$. 
In obvious notations we obtain 
\begin{equation}
\label{37}
R_{AB}=R_{AB}(\frac{1}{\epsilon})+R_{AB}(1)
\end{equation}
in which we have
\begin{equation}
\label{38}
R_{AB}(\frac{1}{\epsilon})=\partial_{a}\Gamma_{AB}^{a}+\Gamma_{ad}^{a} 
\Gamma_{AB}^{d},
\end{equation}
\begin{equation}
\label{39}
R_{AB}(1)=\stackrel{*}{R}_{AB}+\Gamma_{aD}^{D}\Gamma_{AB}^{a}-
\Gamma_{aB}^{C}\Gamma_{AC}^{a}-\Gamma_{BD}^{a}\Gamma_{Aa}^{D}
\end{equation}
with
$\stackrel{*}{R}_{AB}=\partial_{D}\Gamma_{AB}^{D}-\partial_{B}\Gamma_{AD}^{D} 
+\Gamma_{CD}^{D}\Gamma_{AB}^{C}-\Gamma_{BD}^{C}\Gamma_{AC}^{D}$.
  
For the small case indices we get
\begin{equation}
\label{41}
R_{ab}=R_{ab}(\frac{1}{\epsilon^{2}})+R_{ab}(\frac{1}{\epsilon})+R_{ab}(1)
\end{equation}
in which we have 
\begin{equation}
\label{42}
R_{ab}(\frac{1}{\epsilon^{2}})=\stackrel{\wedge}{R}_{ab}
\end{equation}
where $\stackrel{\wedge}{R}_{ab}$ is the Ricci tensor associated with the 
connection (\ref{36}) of the smoothed cone,
\begin{equation}
\label{43}
R_{ab}(\frac{1}{\epsilon})=\Gamma_{dD}^{D}\Gamma_{ab}^{d}
\end{equation}
and 
\begin{equation}
\label{44}
R_{ab}(1)=-\partial_{b}\Gamma_{Da}^{D}-\Gamma_{bD}^{C}\Gamma_{aC}^{D}.
\end{equation}
Finally for the mixed indices we get
\begin{equation}
\label{45}
R_{aB}=R_{aB}(1)=\partial_{A}\Gamma_{aB}^{A}-\partial_{B}\Gamma_{Da}^{D}+\Gamma_{CD}^{C}\Gamma_{aB}^{D}-\Gamma_{BC}^{D}\Gamma_{aD}^{C}.
\end{equation}
The curvature scalar $R$ for $l=l_{\small{0}}$  is given by the equation  
\begin{equation}
\label{46}
R=R(\frac{1}{\epsilon^{2}})+R(\frac{1}{\epsilon})+R(1)
\end{equation}
where
\begin{equation}
\label{47}
R(\frac{1}{\epsilon^{2}})=\stackrel{\wedge}{R}_{ab}g^{ab}
=\stackrel{\wedge}{R},
\end{equation}
\begin{equation}
\label{48}
R(\frac{1}{\epsilon})=\partial_{a}\Gamma_{AB}^{a}g^{AB}+\Gamma_{ad}^{a}
\Gamma_{AB}^{d}g^{AB}+\Gamma_{ab}^{d}\Gamma_{Dd}^{D}g^{ab},
\end{equation}
\begin{equation}
\label{49}
R(1)=\stackrel{*}{R}+\Gamma_{aD}^{D}\Gamma_{AB}^{a}g^{AB}-\Gamma_{aB}^{C}
\Gamma_{AC}^{a}g^{AB}-\Gamma_{BD}^{a}\Gamma_{Aa}^{D}g^{AB}-\partial_{b}
\Gamma_{aD}^{D}g^{ab}-\Gamma_{bD}^{C}\Gamma_{Ca}^{D}g^{ab}
\end{equation}
with
$\stackrel{*}{R}=\stackrel{*}{R}_{AB}g^{AB}$.

We have calculated the above quantities for $l=l_{\small{0}}$ since we need 
these estimations precisely at the boundary in order to obtain the equations 
of motion in the following section. 

\section{Equations of motion of the string}

We now focus attention  on the self-gravitating string of arbitrary shape 
described by 
metric (\ref{20} ) in  the coordinate system $(\tau^{A},\rho^{a})$
 where the function $h$
is specified in Sec. II.  We suppose that the exterior spacetime is a
vacuum solution : 
$R_{\alpha\beta}=0,\; (\alpha=A,a)$ for $l>l_{\small{0}}$.

The energy-momentum tensor $T_{\alpha\beta}$ of the extended string
is the source of the Einstein equations 
\begin{equation}
\label{52}
R_{\alpha\beta}-\frac{1}{2}g_{\alpha\beta}R=8\pi G T_{\alpha\beta}
\quad ,\qquad 0\leq l\leq l_{\small{0}}.
\end{equation} 
As a consequence of the matching conditions adopted, the Ricci tensor must be continuous at the junction $l=l_{\small{0}}$. So, the interior
Einstein equations (\ref{52}) coincide with the vacuum Einstein
equations at $l=l_{0}$ 
\begin{equation}
\label{53}
(R_{\alpha\beta})_{l_{\small{0}}}=0.
\end{equation}
These boundary conditions impose some constraints  on the timelike worldsheet 
swept by the central line of the string. We shall examine (\ref{53}) 
when the parameter $\epsilon$ is arbitrarily small.

According to (\ref{41})-(\ref{44}), the components $(a,b)$ of equations  
(\ref{53}) are written  
\begin{equation}
\label{54}
(R_{ab})_{l_{\small{0}}}=(\Gamma_{dD}^{D})_{l_{\small{0}}}(
\Gamma_{ab}^{d})_{l_{\small{0}}}+(R_{ab}(1))_{l_{\small{0}}}=0,
\end{equation}
equation (\ref{54}) being  obtained by noting from (\ref{42}) that 
$[R_{ab}(\frac{1}{\epsilon^{2}})]_{l_{\small{0}}}=
(\stackrel{\wedge}{R}_{ab})_{l_{\small{0}}}=0$
since the junction $l=l_{0}$ belongs to the cone.

We need to calculate the limit of the connection coefficient $\Gamma_{dD}^{D}$ appearing in equation (\ref{54}). From (\ref{34}), 
(\ref{24}), (\ref{25}) and since $l_{\small{0}}/\epsilon$ is constant 
we obtain 
\begin{equation}
\label{56}
\lim_{\epsilon\rightarrow 0}(\Gamma_{Dd}^{D})_{l_{\small{0}}}=
\lim_{l_{\small{0}}\rightarrow 0}(\Gamma_{Dd}^{D})_{\l_{\small{0}}}=
\gamma^{AB}K_{dAB}=K_{d}
\end{equation}
where $K_{d}$ is the mean curvature of the timelike worldsheet.
On the other hand we know from (\ref{36}) that 
$(\Gamma_{ab}^{d})_{l_{\small{0}}}$ is of order $1/\epsilon$. 
Thus, the limit $\epsilon\rightarrow 0$ of equation (\ref{54}) 
multiplied by $\epsilon$ gives 
\begin{equation}
\label{57}
 F_{ab}^{d}(k^{a})K_{d}=0\quad {\rm with} \quad  k^{a}=
\frac{\rho_{\small{0}}^{a}}{l_{\small{0}}} 
\end{equation}
where
\begin{equation}
\label{58}
 F_{ab}^{d}(k^{a})=\lim_{\epsilon\rightarrow 0}[\epsilon(
\Gamma_{ab}^{d})_{\l_{\small{0}}}] 
\end{equation}
is finite.
Equation (\ref{57}) is independent of $\epsilon$ or $\l_{\small{0}}$ and 
depends only on the azimuthal angle $\phi$ on the boundary or equivalently 
as indicated of the unitary 2-vector $k^{a}$.

In the same way using (\ref{37})-(\ref{39}) along with (\ref{35}),(\ref{36}), 
we obtain the zero limit of $\epsilon$ for the $(A,B)$ components of 
equations (\ref{53}) 
\begin{equation}
\label{59}
F^{b}(k^{a})K_{bAB}=0\quad {\rm with} \quad k^{a}=
\frac{\rho_{\small{0}}^{a}}{l_{\small{0}}} 
\end{equation}
where 
\begin{equation}
\label{60}
F^{b}(k^{a})=\lim_{\epsilon\rightarrow 0}[\epsilon(\partial_{a}g^{ab}+
\Gamma_{ad}^{a}g^{db})_{\l_{\small{0}}}] 
\end{equation}
is finite. By introducing the quantities independent of $\epsilon$ 
\begin{equation}
\label{61}
A=q(\frac{l_{\small{0}}}{\epsilon})\frac{l_{\small{0}}^{2}}{\epsilon^{2}}
\quad {\rm and} \quad B=q^{\prime}(\frac{l_{\small{0}}}{\epsilon})
\frac{l_{\small{0}}^{3}}{\epsilon^{3}},
\end{equation}
after some algebra we can express $F^{b}(k^{a})$ in the following way 
\begin{equation}
\label{62}
F^{b}=\frac{\epsilon}{l_{\small{0}}}\, \frac{1}{1+A}\, [2A+
\frac{1}{2}B] k^{b}.
\end{equation}
Let us note that the mixed components $(a,A)$ of (\ref{53}) give no
constraint.

If the coefficients $F_{ab}^{d}(k^{a})$ and $F^{b}(k^{a})$ are non vanishing, 
which is the general case,  equations (\ref{57}) and (\ref{59}) yield 
respectively 
\begin{equation}
\label{63}
K_{d}=0,
\end{equation}
\begin{equation}
\label{64}
K_{bAB}=0.
\end{equation}
Equation (\ref{64}) means that, at the zero limit of $\epsilon$, the 
worldsheet swept by the central line of the string is totally geodesic. 
This of course implies (\ref{63}). We recover the situation described 
in the litterature \cite{vic,fro,unr,cla,isr2}
 when the string is a self-gravitating singular line.

However we see from (\ref{62}) that if we take the condition
\begin{equation}
\label{65}
B=-4A,
\end{equation}
then one annihilates the extrinsic curvature coefficient and thus constraint 
(\ref{64}) disappears.
Hence we are left with constraint (\ref{63}) which expresses that in the 
zero limit of $\epsilon$ the world sheet is minimal. In other words the 
worldsheet has the same evolution as a Nambu-Goto string. 

We must verify 
however that when condition (\ref{65}) is applied then equation (\ref{57}) 
does not also disappear.
Equations (\ref{57}) are explicitely expressed as
\begin{eqnarray}
\label{66}
\nonumber & & k^{2}[P+Q(k^{1})^{2}]K_{1}+k^{1}[P+Q(k^{2})^{2}]K_{2}=0, \\  
& & -k^{1}Q(k^{2})^{2}K_{1}-k^{2}[2P+Q(k^{2})^{2}]K_{2}=0, \\
\nonumber & & -k^{1}[2P+Q(k^{1})^{2}]K_{1}-k^{2}Q(k^{1})^{2}K_{2}=0
\end{eqnarray}
for all $k^{a}$,
where
$P=B+2A$ and $Q=-B+AB+4A^{2}$.
If $A$ and $B$ verify condition (\ref{65}), then the three equations  
(\ref{66}) are reduced to only one:
$k^{2}K_{1}-k^{1}K_{2}=0$ for all $k^{a}$.
Hence (\ref{63}) is again verified.

By their definition (\ref{61}), the quantities $A$ and $B$ are 
independent of the length   
parameters and depend only on the angle $\alpha$. That is, we can express 
condition (\ref{65}) in terms of the angle $\alpha$ by using (\ref{13}) 
and (\ref{26}). We obtain the simple expression 
\begin{equation}
\label{71}
\sin\alpha\cos\alpha=\frac{l_{\small{0}}}{\epsilon}.
\end{equation}
This relation is a supplementary constraint on $h$ as it will be
explained in the conclusion.

\section{Energy-momentum tensor of the string}
 
The energy-momentum tensor $S_{\alpha\beta}$ of the string is obtained by 
an integration of $T_{\alpha\beta}$ on the section  
$\tau^{A}=$ constant of the string.
\begin{equation}
\label{72}
S_{\alpha\beta}=\int_{l\leq l_{\small{0}}}T_{\alpha\beta}
\sqrt{\stackrel{\wedge}{g}}\, d\rho^{1}\, d\rho^{2}
\end{equation}
where now $\epsilon$ and thus $l_{\small{0}}$ are small but fixed.  

From the Einstein equations (\ref{52}) and the algebraic expressions of the Ricci tensor given in Sec. IV, we can write in obvious notations

\begin{equation}
\label{73}
T_{\alpha\beta}(\tau^A,\rho^a)=T_{\alpha\beta}(\frac{1}{\epsilon^{2}})+T_{\alpha\beta}(\frac{1}{\epsilon})+T_{\alpha\beta}(1)\qquad l\leq l_{\small{0}}.
\end{equation}
We first apply formula (\ref{72}) for the $(A,B)$ components. Only the first term of (\ref{73}) will yield a non null finite integral 
when $\epsilon$ tends to $0$
since the volume element of 
the smoothed cone is of the order $l_{\small{0}}^{2}$.
\begin{equation}
\label{74}
S_{AB}=\int_{l\leq l_{\small{0}}}T_{AB}(\frac{1}{\epsilon^{2}})
\sqrt{\stackrel{\wedge}{g}}\, d\rho^{1}\, d\rho^{2}+O(\epsilon ).
\end{equation}
Now we have
\begin{equation}
\label{75}
T_{AB}(\frac{1}{\epsilon^{2}})=-\frac{1}{16\pi G}g_{AB}
\stackrel{\wedge}{R}=-\frac{1}{8\pi G}g_{AB}K
\end{equation}
where $K$ is the Gauss curvature, therefore
\begin{equation}
\label{76}
S_{AB}=-\frac{1}{8\pi G}
\int_{l\leq l_{\small{0}}}g_{AB}K\sqrt{\stackrel{\wedge}{g}}\, d\rho^{1}\, d\rho^{2}+O(\epsilon).
\end{equation}
Since $l_{\small{0}}$ is small we can also approximate the metric $g_{AB}$ 
by development (\ref{25}).
We finally obtain
\begin{equation}
\label{77}
S_{AB}=-\frac{1}{8\pi G}\gamma_{AB}\int_{l\leq l_{\small{0}}}K\sqrt{\stackrel{\wedge}{g}}\, 
d\rho^{1}d\rho^{2}+O(\epsilon)=-\mu\gamma_{AB}+O(\epsilon)
\end{equation}
where $\mu$ is the linear mass density.
The Gauss-Bonnet formula gives
\begin{equation}
\label{78}
\int_{l\leq l_{0}}K\sqrt{\stackrel{\wedge}{g}}d\rho^{1}d\rho^{2}=2\pi (1-\sin \alpha )
\end{equation}
and consequently $\mu$ is related to $\alpha$ by formula (\ref{9}).
This result is not obvious a priori. It originates specifically in the choice of the form of the metric (\ref{17}) where $f$ is given by (\ref{12}). A general discussion of the problem of obtaining the energy momentum tensor for a concentrate distribution of matter can be found in \cite{ger}.

The integral of the $(a,A)$ components obviously vanishes as $\epsilon^{2}$. For the integral of the $(a,b)$ components the non null finite part vanishes since ${\stackrel{\wedge}{R}}_{ab}-\frac{1}{2}g_{ab}{\stackrel{\wedge}{R}}=0$ for a 2-dimensional manifold. Thus the integral vanishes as $\epsilon$.

Discarding terms in $\epsilon$, $S_{\alpha\beta}$ reduces to $S_{AB}$ given by (\ref{77}) which is the energy-momentum tensor of the Nambu-Goto string.  
This last result which is proved independently of the equations of 
motion is in accordance with them. It can justify 
{\em a posteriori} the  definition of  a 
self-gravitating string as a smooth cone with metric  (\ref{17}) and (\ref{18}).

\section{Conclusion}

In this paper we have investigated the dynamics of a self-gravitating
string in the limit where the thickness $\epsilon$ becomes negligible.
In the generic case the worldsheet swept by the string (more precisely
by the central line of the string) is a totally geodesic surface
(the extrinsic curvature of the worldsheet is null). This result
could have been somehow guessed since the strings defined as singular
lines of conical points have precisely this property 
\cite{vic,fro,unr,cla}. 

However we have found another pleasant possibility : by
imposing relation (\ref{71}) the extrinsic curvature is no more relevant
and only the mean curvature is null. This expresses that the worldsheet is
extremal which is the behaviour of the Nambu-Goto string.
If we suppose that the linear mass density $\mu$ is given, then the
angle $\alpha$ is fixed by equation (\ref{78}). It is easily seen that
relation (\ref{71}) is a constraint on the function $h$; it can
be rewritten
\begin{equation}
\label{90}
h(\frac{l_{0}}{\epsilon})h'(\frac{l_{0}}{\epsilon})=\frac{l_{0}}{\epsilon}.
\end{equation}
The above equation is an algebraic relation taken at the fixed point $l_{0}/\epsilon$ which is added to the matching conditions (\ref{13}) and (\ref{16}). It is easily shown that such a function $h$ can be found in a polynomial form. The simplest solution is an odd polynomial (cf. (4)) of order seven :
\[
 h(\frac{l_{0}}{\epsilon})=\frac{l_{0}}{\epsilon}+b_{3}(\frac{l_{0}}{\epsilon})^3+b_{5}(\frac{l_{0}}{\epsilon})^5+b_{7}(\frac{l_{0}}{\epsilon})^7.
\]
The four unknown quantities $l_{0}/\epsilon$, $b_3$, $b_5$ and $b_7$ can be determined as functions of the angle $\alpha$ (or equivalently as a function of the linear mass density $\mu$ of the string) by the four equations (\ref{13}), (\ref{16}) and (\ref{90}). Of course by taking a polynomial of larger order one can get an infinity of solutions. It would probably be interesting to investigate the physical meaning of the constraint (\ref{90}) on the matter of the string.

By using the method described in Sec. IV, we can evaluate the Riemann tensor $(R_{\alpha\beta\gamma\delta})_{l_{\small{0}}}$ at the junction $l=l_{0}$ 
or equivalently the Weyl tensor since the Ricci tensor (\ref{53}) vanishes. This gives the magnitude of the gravitational field near the thin cosmic string. In the generic case where the extrinsic curvature $K_{aBC}$ vanishes, the Riemann tensor is bounded as the radius $l_{0}$ becomes arbitrarily small. On the contrary when only the mean curvature $K_a$ is null, the Riemann tensor components $(R_{aBCd})_{l_{\small{0}}}$  blow up as $1/\epsilon$, {\it i.e.} $1/l_{0}$,  when the radius $l_{0}$ tends to zero. We see that the gravitational aspect is completely different in the latter case. Let us add that if we require that the Riemann tensor remains bounded then we necessarily fall in the generic case. 
 

\newpage
\begin{thebibliography}{99}

\bibitem{she} A. Vilenkin and E.P.S. Shellard,
{\em Cosmic strings and other topological defects} (Cambridge University
Press, Cambridge, 1994).

\bibitem{isr1} W. Israel, Phys. Rev. D {\bf 15}, 935 (1977).

\bibitem{vil} A. Vilenkin, Phys. Rev. D {\bf 23}, 852 (1981).

\bibitem{vic} J.A.G. Vickers, Class. Quantum Grav. {\bf 4}, 1 (1987).

\bibitem{fro} V.P. Frolov, W. Israel, and W.G. Unruh, Phys. Rev. D {\bf 39},
1084 (1989).

\bibitem{unr} W.G. Unruh, G. Hayward, W. Israel, and D. McManus, 
Phys. Rev. Lett. {\bf 62}, 2897 (1989).

\bibitem{cla} C.J.S. Clarke, G.F.R. Ellis, and J.A. Vickers, 
Class. Quantum Grav. {\bf 7}, 1 (1990). 

\bibitem{isr2} W. Israel, in 
{\it Journ\'{e}es Relativistes 1989} (Tours) edited by C. Barrab\`es and
B. Boisseau, Annales de Physique {\bf 14}, Colloque 
${\rm n}^{\rm o}$ 1 (1989). 
\bibitem{mar} L. Marder, Proc. Roy. Soc. London A {\bf 252}, 45 (1959).
\bibitem{got} J.R. Gott III, Astrophys. J. {\bf 288}, 422 (1985).
\bibitem{his} W.A. Hiscock, Phys. Rev. D {\bf 31}, 3288 (1985).
\bibitem{lin} B. Linet, Gen. Rel. Grav. {\bf 17}, 1109 (1985).

\bibitem{boi} B. Boisseau, and P.S. Letelier, Phys. Rev. D {\bf 46}, 1721 
(1992).
\bibitem{ger} R. Geroch and J. Traschen, Phys. Rev. D {\bf 36}, 1017 (1987).
\end{thebibliography}
\end{document}